\documentclass[amsmath,amssymb,preprint]{elsarticle}

\usepackage{amssymb}
\usepackage{amsmath}
\usepackage{graphicx}
\usepackage{array}
\usepackage{color}

\newcommand{\ket}[1]{\left\vert #1 \right>}
\newcommand{\pare}[1]{\left( #1 \right)}

\begin{document}

\title{Two-Photon Absorption Spectroscopy using Intense Phase-Chirped Entangled
Beams}

\author{Ji\v{r}\'i Svozil\'{i}k}

\address{Quantum  Optics  Laboratory, Universidad  de  los  Andes,  A.A.
    4976,  Bogot\'a  D.C.,  Colombia}

\address{Joint Laboratory of Optics of Palack\'y University and
    Institute of Physics of CAS, 17.~listopadu~12, 771 46 Olomouc, Czech Republic}

\author{Jan Pe\v{r}ina Jr.}

\address{Joint Laboratory of Optics of Palack\'y University and
    Institute of Physics of CAS, 17.~listopadu~12, 771 46 Olomouc, Czech Republic}

\author{Roberto de J. Le\'{o}n-Montiel}
\address{Instituto de Ciencias Nucleares, Universidad Nacional
    Aut\'{o}noma de M\'{e}xico, Apartado Postal 70-543, 04510 Cd. Mx.,
    M\'{e}xico}

%\pacs{42.50.Ct, 42.50.Hz}

%
%\mail{jiri.svozilik@gmail.com}

%\ocis{(270.4180)Multiphoton processes; (300.6410) Spectroscopy, multiphoton}

\begin{abstract}
We numerically analyze the use of intense entangled twin beams for
ultra-sensitive spectroscopic measurements in chemical and
biological systems. The examined scheme makes use of intense
frequency-modulated (chirped) entangled beams to successfully
extract information about the intermediate material states that
contribute to the two-photon excitation of an absorbing medium.
Robustness of the presented method is examined with respect to the
applied intervals of the frequency chirp.
\end{abstract}

\maketitle

\section{Introduction}

Quantum entanglement, which lies at the heart of modern quantum spectroscopy
\cite{Saleh1998Entangled,Leon2013Role,schlawin2017entangled} has been
recognized as a powerful
resource for the development of novel methods and applications in various fields
of research, including quantum cryptography \cite{Ekert1991Quantum}, quantum
computing \cite{Nielsen2010Quantum}, and quantum metrology
\cite{Giovannetti2004Quantum}. In particular, regarding the latter, the use of
entangled light in two-photon absorption (TPA) spectroscopy has already
received a great deal of attention
\cite{Lee2006Entangled,guzman2010,Roslyak2009,Schlawin2012,Raymer2013,dorfman2016,villabona2017entangled}
 because of the unique phenomena that arise in the interaction of entangled
photon pairs with matter. As examples, we mention linear scaling of the
TPA rates on the photon flux \cite{Javanainen1990}, two-photon-induced
transparency \cite{Fei1997Entanglement}, the ability to select different states
in complex biological aggregates \cite{Schlawin2013Suppression}, and the control
of entanglement in matter \cite{Shapiro2011,Shapiro_book}. Indeed, the
prediction and observation of these fascinating effects can be understood as a
direct consequence of the dependence of the TPA signal on the properties
of quantum light that interacts with the sample
\cite{Perina1991Quantum,Dayan2007Theory,oka2010efficient}.

Among different techniques proposed over the years,
entangled-photon based spectroscopy techniques such as the
virtual-state (VSS) and pump-probe (PPS) spectroscopies
\cite{Saleh1998Entangled,Perina1998Multiphoton,Leon2013Role,schlawin2016pump,schlawin2017entangled}
have proved to be a unique tool for extracting information about
the intermediate levels \cite{Shore1979,SakuraiBook}, that
contribute to the two-photon excitation of an absorbing medium. In
these techniques, intermediate transitions, a signature of the
medium, are experimentally revealed by introducing a time delay
between photons comprising a state entangled in frequencies.

Regardless of some experimental issues expressed in the original
proposal of VSS, it can be considered as a new route towards novel
applications in ultra-sensitive detection \cite{Lee2007entangled}.
In this contribution, we propose an experimental scheme that can
successfully overcome two issues. Generally speaking, quantum
based spectroscopic methods are usually experimentally challenging
due to the low TPA rates and required weak sources of photon
pairs. This first issue is addressed by considering that the
medium interacts with a twin beam composed by a larger number of
entangled photon pairs and generated by means of intense
parametric down-conversion (PDC)
\cite{Svozilik2009Properties,Svozilik2011Generation,Valles2013Generation}.
Indeed, properties of such twin beams have been commonly
investigated \cite{Machulka2013,Allevi2014Coherence,Perez2014}
and, more importantly, it has been shown that strong quantum
frequency correlations inside the twin beam persist even at these
higher photon-flux conditions as shown in
\cite{Schlawin2013Photon} and also later in
\cite{haderka2015spatial,Perina2015Coherence,PerinaJr2016a}. {
A 3D model of an intense twin beam developed in \cite{PerinaJr2016}
has revealed that typically tens of independent dual spectral
Schmidt modes are found in such intense twin beams in quantum
superpositions. For ultrashort pump pulses, these Schmidt modes
are delocalized in the whole twin-beam spectral range which
guarantees the performance of the VSS in the whole used spectral
region. Though the weights of individual members in these quantum
superposition change with the increasing twin-beam intensity,
there still remains sufficient numbers of comparably populated
members that keep quantum character of the twin-beam state that
allow for the VSS.} The use of such twin beams in the quantum
spectroscopy has been recently discussed in \cite{Svozilik2018}.

The second issue comes with the necessity of averaging over
several realizations differing in temporal correlations
(entanglement) between the photons obtained by either changing the
width of the pump pulse (in Type-I nonlinear process) or the
length of the nonlinear crystal (Type-II) in each realization
\cite{Saleh1998Entangled}, which can be experimentally cumbersome.
This approach is required to correctly identify the transition
frequencies in a detected signal from their combinations. { We
note that the detected signal encompasses contributions from all
possible quantum paths that lead to the absorption of two photons
with the participation of two arbitrary intermediate levels.
However, only the paths at which the same intermediate level is
used give us the looked-for energies of the intermediate levels.}
Alternatively to the discussed averaging approaches, we may vary
and control the properties of entangled photon pairs by modifying
the spatial shape of the pump beam
\cite{Carrasco2004Spatial,Valencia2007Shaping}. To address this
issue, we make use of a technique in which the spectral
entanglement between photons is modified by frequency variations
in the spectrum of the signal beam. Such approach has been
extensively studied in
Refs.~\cite{meshulach1998coherent,dayan2004two,joseph2015entanglement}.
Because typical correlation times of down-converted photons are
found in the femtosecond domain, one needs to make sure that any
frequency modulation of photons is done in the ultrafast regime.
Such methods already exist, in particular those based on the use
of spatial light modulators, which have shown to be extremely
efficient for femtosecond pulse shaping [43].

\begin{figure}[t]
 \centering
 \includegraphics[width=9cm]{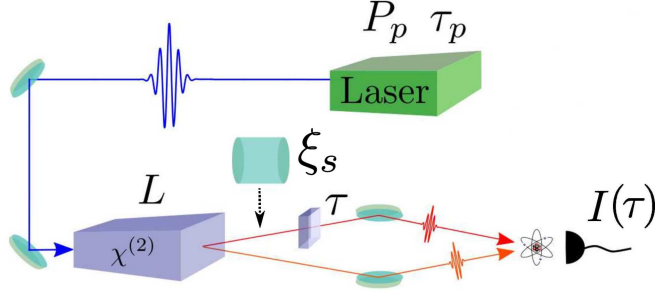}
 \caption{Proposed experimental setup. A pump pulse, with time duration $\tau_{p}$
 and power $P_p$, interacts with a nonlinear crystal of length $L$, which serves as a source of twin beams composed
 of many photon pairs. The generated photons are spatially separated and a time delay $\tau$ between them is
 introduced. Information about virtual transitions is then obtained by monitoring the two-photon absorption rate as
 a function of delay $\tau$. In order to obtain the relative variances of resolved
 spectral peaks, the signal-beam
 spectrum is modulated by a phase modulator generating a frequency chirp
 $\xi_s$.}
\label{Fig:1}
 \end{figure}

Here we show that this technique can also be used to extract
information about the energy-level structure of the sample by
computing the relative variance of several TPA signals obtained
for different signal frequency chirps, as depicted in
Fig.~\ref{Fig:1}. Robustness of the proposed method, which is an
important practical issue, is also discussed later.

Our proposed scheme might open a new avenue towards the
experimental implementation of quantum spectroscopy by overcoming
the low photon-flux issue, while maintaining the most important
feature of this technique which lies in the fact that, unlike
other commonly used TPA spectroscopy techniques that require
sophisticated tunable sources, it can be implemented by carrying
out pulsed or continuous-wave absorption measurements without
changing the wavelength of the source
\cite{Saleh1998Entangled,Kojima2004Entangled,Leon2013Role}. {
The use of spectrally broad-band photon-pair fields, that may
cover even 2-3 hundreds of nm in the visible spectral range,
together with the fact that VSS scans the whole spectral range in
each measurement repetition make the VSS in principle superior
above the other conventionally used spectroscopic methods with
tunable light sources. We note that the effect of scanning the
whole spectral range in each experimental repetition is reached
also when two ultrashort pulses with exactly opposed chirps are
applied to observe two-photon absorption, as suggested in
\cite{Schlawin2013Suppression}. However, practical realization of
this method would require perfect synchronization of both chirps
and moreover the covered spectral range, that is derived from the
pulses spectral widths, would be quite narrow.}

\section{Theoretical background}

Let us consider the interaction of a medium with a twin beam,
described by the interaction Hamiltonian
$\hat{H}_{I}\pare{t}=\hat{\mu}\pare{t}\hat{E}^{\pare{+}}\pare{t}$,
where $\hat{\mu}\pare{t}$ is the medium dipole-moment operator and
$\hat{E}^{\pare{+}}\pare{t} = \hat{E}_{s}^{\pare{+}}\pare{t+\tau}
+ \hat{E}_{i}^{\pare{+}}\pare{t}$ is the overall
positive-frequency electric-field operator, with subscripts $s$
and $i$ denoting the signal and idler fields that constitute the
twin beam. Notice that we have included a time delay $\tau$
between the fields, which is needed for implementing the pump-probe
technique.

We assume that the medium is initially in its ground state $\ket{g}$ (with energy $\epsilon_{g}$). Then, the
probability that the medium is excited into its final state $\ket{f}$ (with energy $\epsilon_{f}$), through a TPA
process, is obtained by means of second-order time-dependent perturbation theory in the form
\cite{Perina1998Multiphoton,Shen1984Principles,Mukamel1999Principles}:
\begin{eqnarray}
\begin{split}
P_{g\rightarrow f}&=\int_{-\infty}^{\infty}
dt_{2}\int_{-\infty}^{t_{2}}dt_{1}\int_{-\infty}^{\infty}dt'_{2}\int_{-\infty}^{t'_{2}}dt'_{1}
 M^{*}\left(t_2,t_1\right)\nonumber\\
&\times G^{(2)}\pare{t_{2},t_{1};t'_{2},t'_{1}}M\left(t'_2,t'_1\right),
\label{Eq:1}
\end{split}
\end{eqnarray}
where $ M $ stands for the two-point temporal dipole-moment
correlation function,
\begin{equation}
 M(t_2,t_1)=\sum_j \frac{\mu_{fj}\mu_{jg}}{\hbar^{2}}
  \exp\left[i\left(\epsilon_f-\epsilon_j\right)t_2+i\left(\epsilon_j-\epsilon_g\right)t_1
  \right],
\label{EqM}
\end{equation}
$ G^{(2)} $ denotes the optical-field fourth-order
\begin{equation}
\begin{split}
G^{(2)}\pare{t_{2},t_{1};t'_{2},t'_{1}} = &\langle \hat{E}^{(-)}\left(t_2\right)\hat{E}^{(-)}\left(t_1\right)\hat{E}^{(+)}\left(t'_2\right)\hat{E}^{(+)}\left(t'_1\right)\rangle ,
\end{split}
\label{EqG}
\end{equation}
and $ \hbar $ stands for the reduced Planck constant.

Equation (\ref{EqM}) describes the response of the medium to the
applied twin beam and $\mu_{fj}$ and $\mu_{jg}$ are the transition
matrix elements of the dipole-moment operator between the states
specified in the subscript. Notice that the excitation of the
medium occurs through intermediate, fast decaying, states
$\ket{j}$, with complex energy eigenvalues $\epsilon_{j} =
\tilde{\epsilon}_{j} - i\kappa_{j}$; $\tilde{\epsilon}_{j}$ is the
energy eigenvalue of the $j$th intermediate level and $\kappa_{j}$
gives its natural line-width.

On the other hand, Eq.~(\ref{EqG}), giving the fourth-order field
correlation function \cite{glauber1965,mandel1965}, characterizes
statistical properties of the interacting optical fields. In what
follows, we concentrate on the description of the quantum fields
comprising the twin beam, whose temporal (as well as spectral)
correlations are controlled by frequency variations in the
spectrum of the signal field. { We note that these correlations
are by definition quantum as they describe the properties of
spectral quantum superpositions of the Schmidt dual modes
comprising the twin beam (see below).} The intense twin beams
composed of many photon pairs can be conveniently described by
using the
 two-step procedure
\cite{Perina2015Coherence,PerinaJr2016,Wasilewski2006Pulsed,Schlawin2013Photon,Svozilik2018}.
Photon fields occurring in Eq.~(\ref{EqG}) are described by the
field operators \cite{Huttner1990}:
\begin{eqnarray}
\hat{E}^{(+)}_{j}\left(z,t\right)&=&\int d\omega_j
\sqrt{\frac{\hbar\omega_j}{2\pi
        cA_{q}n_j}}\hat{a}_j\left(z,\omega_j\right)\nonumber\\
&\times& \exp\left[-i\pare{\omega_jt-k_{zj}\left(\omega_j\right)z}\right],
\hspace{3mm} j=s,i;
\label{Eq:2}
\end{eqnarray}
$n_j$ is the index of refraction of field $ j $, and
$\hat{a}_j\left(z,\omega_j\right)$ stands for the photon
annihilation operator of a mode with frequency $\omega_{j}$. The corresponding
longitudinal wave-vector is denoted
as $k_{zj}$. $A_{q}$ is the quantization area.

The standard procedure for obtaining the quantum state of photon
twin beams, that is based on the first-order perturbation solution
of the Schr\"odinger equation, results in the following spectrally
entangled two-photon state $ |\Phi\rangle $:
\begin{eqnarray}
|\Phi\rangle&=&\int_{-\infty}^{\infty}d\omega_s\int_{-\infty}^{\infty}d\omega_i\Phi(\omega_s,\omega_i)\hat{a}^{\dagger}_{s}
(\omega_s)\hat{a}^{\dagger}_{i}(\omega_i)|{\rm vac}\rangle.\nonumber
\label{Eq:5}
\end{eqnarray}
The two-photon spectral amplitude
$\Phi\left(\omega_s,\omega_i\right)$ introduced in
Eq.~(\ref{Eq:5}) is equal to:
\begin{eqnarray}
\Phi(\omega_s,\omega_i)&=&C_\Phi\sqrt{\omega_s\omega_i}\exp\left[-\frac{\tau^2_p}{4}
\left(\omega_s+\omega_i-\omega^{0}_{p}\right)^2 \right]\\
& & \hspace{-12mm}  \times\int^{0}_{-L}dz\exp\left(
i\left[k_{zp}(\omega_s+\omega_i)-k_{zs}(\omega_s)-k_{zi}(\omega_i)\right]z\right).\nonumber
\label{Eq:6}
\end{eqnarray}
where the constant $C_\Phi=\frac{2i\chi^{(2)}\eta_p}{\sqrt{2\pi
n_{p}n_{s}n_{i}}}\sqrt{\frac{\tau_p}{\sqrt{2\pi}}}$. The
pump-pulse duration is denoted as $\tau_p$. The parameter
$\eta_p=\sqrt{P_p/\epsilon_0cfn_p}$ quantifies the pump amplitude,
which depends on pump power $P_p$, pump repetition rate $f$,
medium index of refraction $n_p$, and the speed of light $c$. The
central frequency of the pump field is denoted as
$\omega_{p}^{0}$. The nonlinear medium of length L is
characterized via the nonlinear susceptibility $\xi^{(2)}$. The
Schmidt decomposition of the normalized two-photon spectral
amplitude $\tilde{\Phi} $, $ \tilde{\Phi}=\Phi / \mathcal{N}$ and
$\mathcal{N}^2=\int d\omega_s\int
d\omega_i|\Phi\left(\omega_s,\omega_i\right)|^2$, reveals pairs of
spectral modes \cite{Law2000Continous,Law2004Analysis} obeying the
relation:
\begin{equation}
\tilde{\Phi}\left(\omega_s,\omega_i\right)=\sum_{g=1}^{\infty}\lambda_{g}
f^{*}_{s,g}\left(\omega_s\right)f^{*}_{i,g}\left(\omega_i\right).
\label{Eq:8}
\end{equation}
Here $\left\{\lambda_g\right\}_{g=1}^{\infty}$ is the set of
singular values belonging to singular vectors
$\left\{f_{s,g}(\omega_s)\right\}_{g=1}^{\infty}$ and
$\left\{f_{i,g}(\omega_i)\right\}_{g=1}^{\infty}$. Then the
creation operators of Schmidt modes $\hat{a}^{\dagger}_{s,g}$
[$\hat{a}^{\dagger}_{i,g}$] for the signal (idler) eigenfunction
$f_{s,g}\left(\omega_s\right)$ [$f_{i,g}\left(\omega_i\right)$]
are defined by the formula:
\begin{equation}
\hat{a}^{\dagger}_{j,g}=\int_{-\infty}^{\infty}f_{j,g}^*\left(\omega_j\right)\hat{a}^{\dagger}_j\left(\omega_j\right),
\hspace{3mm} j=s,i.
\end{equation}
The inverse relation attains the form:
\begin{equation}
\hat{a}_j^{\dagger}\left(\omega_j\right)=\sum_{g=1}^{\infty}f_{j,g}
\left(\omega_j\right)\hat{a}_{j,g}^{\dagger}. \label{Eq:10}
\end{equation}
These Schmidt modes enter to the Heisenberg equations
\cite{Schlawin2013Photon,Perina2015Coherence} to obtain the
solutions for monochromatic fields:
\begin{equation}
\hat{a}_s(\omega_s,L)=\sum_{g=1}^{\infty}f_{s,g}^{*}(\omega_s)
\left[u_g\hat{a}_{s,g}(0)+v_g\hat{a}^{\dagger}_{i,g}(0)\right].
\label{Eq:14}
\end{equation}
here $u_g=\cosh(\mathcal{N}\lambda_g)$ and
$v_g=\sinh(\mathcal{N}\lambda_g)$. Substitution of
Eq.~(\ref{Eq:14}) into Eq~(\ref{Eq:2}) giving the signal and idler
electric-field amplitudes $\hat{E}_j^{(+)}\left(z=L,t\right)$,
$j=s,i $, allows us to evaluate the optical-field fourth-order
correlation function $G^{(2)}(t_1,t_2,t'_1,t'_2)$ defined in
Eq.~(\ref{EqG}).

To identify the correct intermediate levels among all spectral
peaks, we introduce a quadratic frequency chirp, controlled by the
parameter $\xi_{s}$, to the signal-photon path (or, equivalently,
to the idler-photon path). This is accomplished by transforming
the signal-photon operators given by Eq.(\ref{Eq:14}) in the
following way
\begin{equation}
\hat{a}_s\left(z,\omega_s\right)\rightarrow
 \exp\left[i\xi_{s}\left(\omega_s-\omega_{s0}\right)^2\right]
 \hat{a}_s\left(z,\omega_s\right).
\label{decomp2}
\end{equation}

Finally, the mean photon-pair number $ N_j $ of the twin beam is
determined by $N_{j}=\int d\omega_j \langle\hat{a}^{\dagger}_{j}
\left(\omega_j\right)\hat{a}_{j}
\left(\omega_j\right)\rangle=\sum_{g=1}^{\infty}\langle\hat{a}^{\dagger}_{j,g}
\hat{a}_{j,g}\rangle=\sum_{g=1}^{\infty}|v_g|^2$. This means that
in order to have intense twin beams, the condition $ N_{j} =
\sum_{g=1}^{\infty}|v_g|^2 \gg 1$ has to be satisfied.

\section{Numerical Results}

\begin{figure}[t]
    \centering
    \includegraphics[width=12cm]{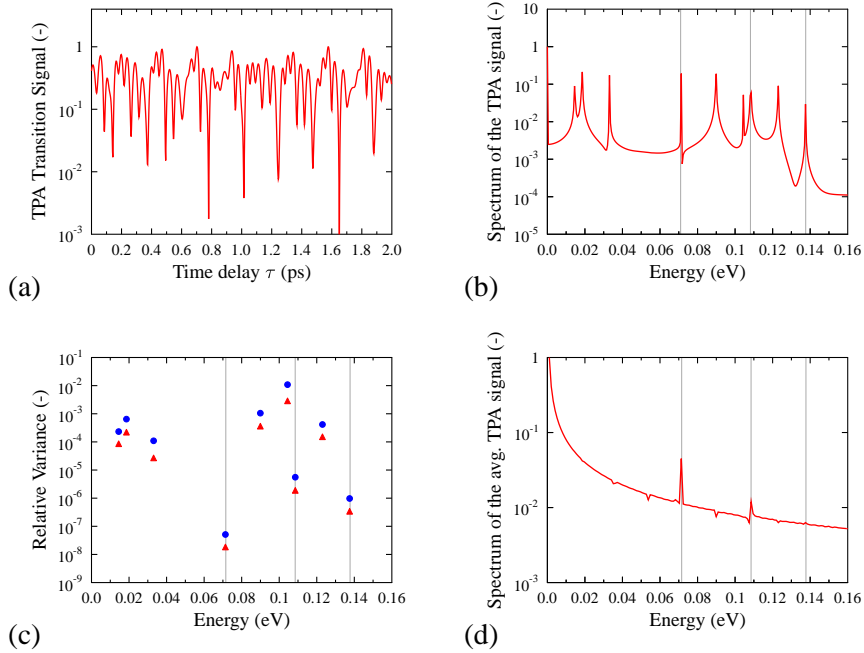}
    \caption{ (a) TPA transition signal as a function of the delay $\tau$
        between fields carrying $N_{j}\approx 100 $ photons. Spectral
        decomposition [Eq.~(\ref{Eq:8})] of the twin beams was done using 500
        singular vectors. (b) Fourier transform of the TPA transition
        probability shown in (a). (c) Relative variance of
        transition-probability  fluctuations of the peaks resolved in the TPA
        spectra obtained for 20 different values of the signal-frequency chirp
        $\xi_s\in\langle0.0,9.5\rangle$ fs$^2$ (blue
        circles) and 40 values $\xi_s\in\langle0.0,3.9\rangle$ fs$^2$ (red
        triangles). The crystal has the length $L=0.801$~mm. (d) Spectrum
        of the TPA transition probability averaged over an ensemble of 100
        crystals
        of different lengths $L\in\langle20,22\rangle$~mm. The vertical grey
        lines indicate the doubled relative energies $|2\epsilon_j-\epsilon_f|$
        of the intermediate levels.}
    \label{Fig:2}
\end{figure}

With the aim of demonstrating the capability of the proposed
scheme for extracting information about the energy level structure
of the medium, we consider a model system whose two-photon
transitions take place via three intermediate states with
randomly-chosen energies $\epsilon_{j}=\{1.586, 1.604, 1.619\}$
~eV and the doubly-excited state satisfying the condition
$\epsilon_{f} - \epsilon_{g}=3.0996$~eV. For the simplicity, we
set all transition dipole moments $\mu_{fk}$ and $\mu_{kg}$ such
$\mu_{fk}\mu_{kg}/\hbar^2=1$ $\mathrm{m^2.V^{-2}.s^{-2}}$. In
addition, we assume the intermediate-state lifetimes much larger
than the twin-beam time duration as well as the relative delay $
\tau $ \cite{Saleh1998Entangled,Kojima2004Entangled,Leon2013Role}.
As the twin-beam source, we consider a near-to-collinear PDC with
Type-II interaction, where a nonlinear crystal, of length
$L=0.801$~mm and with inverse group velocities $G_p=5.4$~ps/m,
$G_s=5.2\,$~ps/m, $G_i=5.6$~ps/m of the pump, signal, and idler
photons, is illuminated by a laser pulse of wavelength
$\lambda_{p}=0.4$~$\mu$m with time duration $\tau_p=1$~ps. For the
sake of simplicity, we assume a spectrally degenerate twin beam,
i.e. $\lambda_{s}=\lambda_{i}=\lambda_p/2$.

Figure~\ref{Fig:2}(a) shows the TPA transition signal
[Eq.~(\ref{Eq:1})] as a function of the delay between the pulses,
each of which carrying $N_{j} \approx 100$ photons. The TPA
transition signal is normalized with respect to its maximal value
in the whole range of $\tau$.  Notice the non-monotonic behaviour
of the TPA transition probability. This phenomenon, related to the
entanglement-induced two-photon transparency
\cite{Fei1997Entanglement}, arises from the
quantum interference between different interaction pathways,
shown in Figure 3, through which two-photon excitation of the
medium can occur. To retrieve the spectroscopic information
present in the TPA signal, one might feel tempted to directly
Fourier transform Eq.~(\ref{Eq:1}). However, as pointed out in
Ref.~\cite{Saleh1998Entangled}, the TPA signal as a function of
delay $\tau$ contains spectral components at various intermediate
frequencies that hinder the clear identification of intermediate
transitions [see Figures ~\ref{Fig:2}(a,b) and ~\ref{Fig:3} ]. To
address this issue, we introduce a new method that benefits from
different strengths of transition probability fluctuations
characterizing the sought and undesired peaks, which originate
from the twin-beam spectral changes induced by the
signal-frequency chirp $\xi_{s}$. These fluctuations are
quantified by the relative variance $ R_n $ defined as:
\begin{equation}
R_{n}=\frac{\rm{var}_{\xi_s}\left[P_{n}\right]}{M_{\xi_s}\left[P_{n}\right]^2},
\end{equation}
where $P_{n}$ is the transition probability of the $n$th spectral
peak, and $M_{\xi_s}\left[P_{n}\right]$ its mean value.
Figure~\ref{Fig:2}(c) shows the relative variance of the resolved
peaks present in the spectrum of the TPA signal
[Fig.~\ref{Fig:2}(b)]. It is clearly visible that the peaks with
the lowest fluctuations (marked with vertical gray lines, notice
the vertical logarithmic scale) are located at the sought energies
$|2\epsilon_j-\epsilon_f|$. This allows us to immediately identify
the intermediate states contributing to the two-photon excitation
of the medium. The success of this technique resides in the fact
that the signal-frequency chirp does not disturb probabilities of
these peaks, whereas the probabilities of the rest of the peaks,
that are undesired, strongly fluctuate around their mean values.

\begin{figure}[t]   % figure 3
    \centering
    \includegraphics[width=11cm]{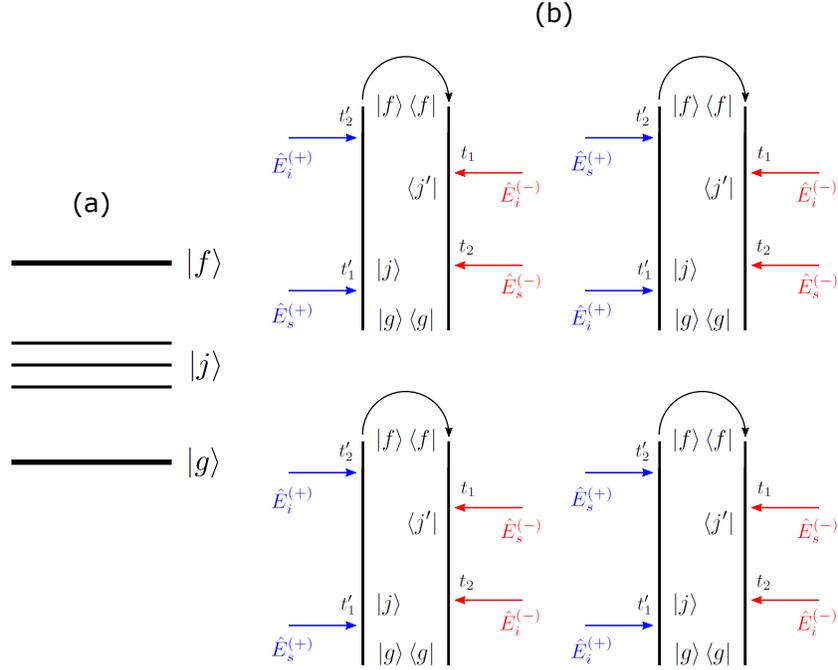}
    \caption{(a) Scheme of energy levels used in our simulations.
    (b) {\bf Closed} time path loop diagrams \cite{Roslyak2009,Roslyak2009-1},
    representing the entangled photon-matter quantum {\bf interaction} pathways
    that contribute to the non-monotonic behaviour of the TPA signal. The 
    time-ordered photon-absorption events can be seen by moving clockwise along 
    the loops.}
    \label{Fig:Pathways}
\end{figure}

To compare our approach with the original proposal
\cite{Saleh1998Entangled}, we perform in parallel an average of
the TPA signal over different crystal lengths. The obtained
spectrum of the averaged signal is plotted in Fig.~\ref{Fig:2}(d).
In the spectrum, the positions of the peaks that emerge from the
Fourier transform of the averaged TPA signal exactly match those
identified in our approach by the lowest available fluctuations.
This demonstrates that our spectroscopic method can successfully
retrieve information about the intermediate-state energies.
Importantly, in contrast to the original implementation of VSS, the average is
done by simply changing the chirp introduced into the signal beam.

\section{Robustness of the proposed scheme}

\begin{figure}[h]
    \centering
    \includegraphics[width=12cm]{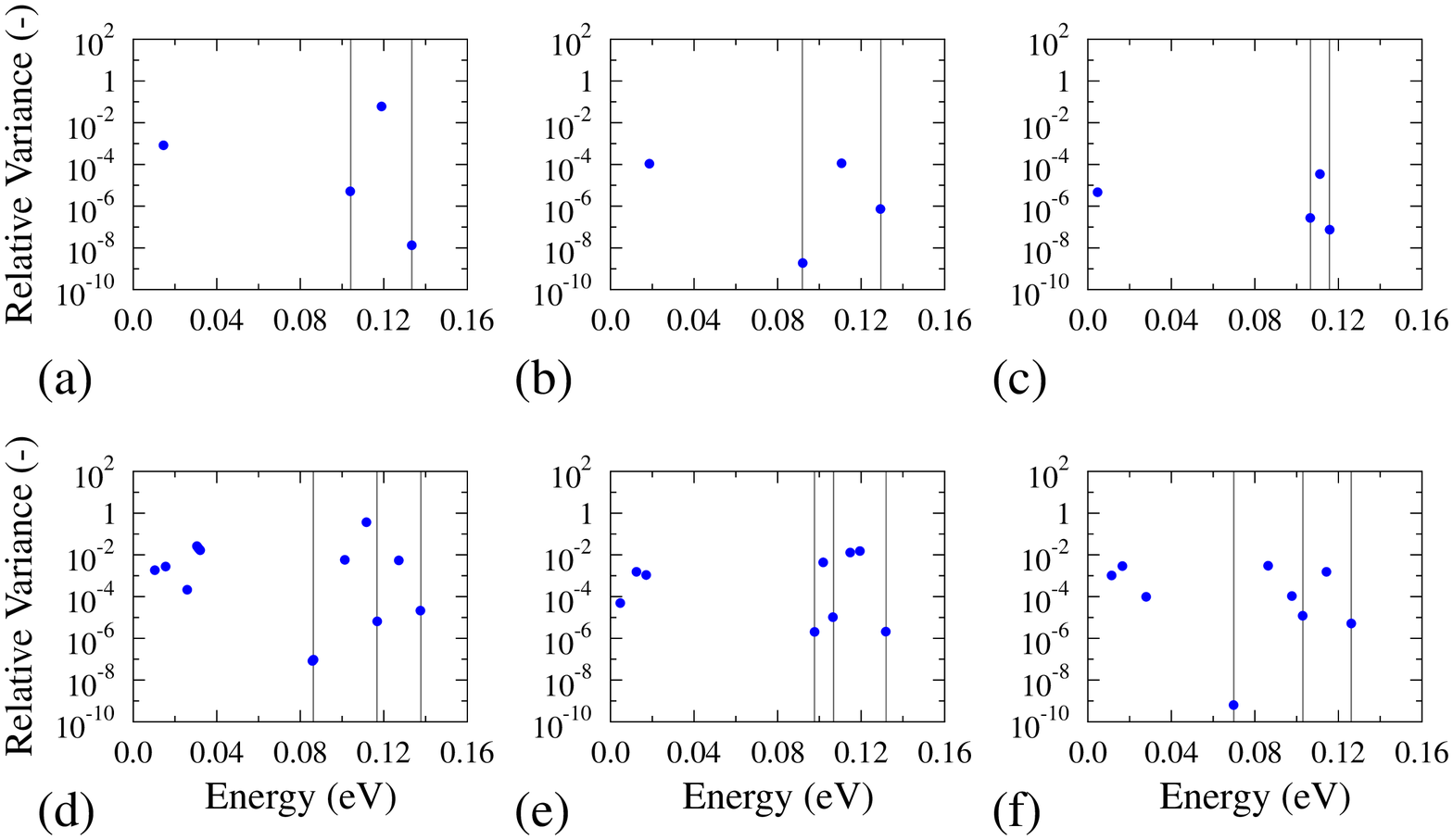}
    \caption{ Relative variances originating from (a-c) 2-intermediate levels
    and (d-f) 3
        intermediate levels considering 20 steps of the chirp parameter $\xi_s$ in the
        interval $\langle0,19\rangle$ fs$^2$. The used crystal
        length equals (a)
        $0.8\,\mathrm{\mu m}$ , (b) $0.55\,\mathrm{\mu m}$, (c)
        $0.65\,\mathrm{\mu m}$, (d) $0.8\,\mathrm{\mu m}$, (e)
        $0.8\,\mathrm{\mu m}$, and (f) $0.716\,\mathrm{\mu m}$. }
    \label{Fig:3}
\end{figure}

Since a full-scale proof of the presented method is numerically
highly demanding, we show the application of our approach on
additional examples of model systems having 2 and 3 intermediate
levels with random energies. For each configuration, a suitable
crystal length has been used and subsequently calculations of TPA
signals for varying $\xi_s$ have been performed. From the obtained
spectra of TPA signals we calculate the relative variances. In
Fig.~\ref{Fig:3} we show the results for different
intermediate-level configurations. Plots (a-c) of the first row of
Fig.~\ref{Fig:3} correspond to the case with 2 intermediate
levels, whereas the plots (d-f) in the second row show the cases
with 3 intermediate levels. In all presented cases, the amount of
signal photons have been maintained in the regime where the
contributions originating in the quantum field entanglement
prevail.

Notice that the results presented in Fig.~\ref{Fig:3} show that
the energy levels of the sample appear as those with the minimal
relative variances, thus providing a simple way to characterize
the electronic-level structure of an unknown sample.

\section{Conclusion}

A new scheme for the implementation of quantum entangled
spectroscopy has been suggested and analyzed. The proposed scheme
makes use of intense twin beams with the spectral chirp introduced
to the signal beam. With our technique, we were able to
successfully retrieve the information about the energy-level
structure of an absorbing medium by statistically analyzing
two-photon transition probabilities obtained for different mutual
time delays and signal-frequency chirps, using just a single
nonlinear crystal.

\section{Acknowledgement}

This work was supported by the projects No.~18-22102S (J.P.)
and 17-23005Y (J.S.) of the Czech Science Foundation. J.S. also
acknowledges the Faculty of Science of Universidad de los Andes. R.J.L.M. thankfully
acknowledges financial support by CONACYT under the project CB-2016-01/284372, and by DGAPA-UNAM under the project UNAM-PAPIIT IA100718.

\bibliographystyle{elsarticle-num}
\bibliography{TPA_Svozilik}

\end{document}